# Trial of an AI: Empowering people to explore law and science challenges

*Arthur Gaudron, Center for Robotics*
*MINES ParisTech, PSL Research University*
*60 Bd. St Michel 75006 Paris, France*
*arthur.gaudron@mines-paristech.fr*
*+33140519127*

Artificial Intelligence represents many things: a new market to conquer or a quality label for tech companies, a threat for traditional industries, a menace for democracy, or a blessing for our busy everyday life. The press abounds in examples illustrating these aspects, but one should draw not hasty and premature conclusions. The first successes in AI have been a surprise for society at large – including researchers in the field. Today, after the initial stupefaction, we have examples of the system reactions: traditional companies are heavily investing in AI, social platforms are monitored during elections, data collection is more and more regulated, etc. The resilience of an organization (i.e. its capacity to resist to a shock) relies deeply on the perception of its environment. Future problems have to be anticipated, while unforeseen events occurring have to be quickly identified in order to be mitigated as fast as possible. The author states that this clear perception starts with a common definition of AI in terms of capacities and limits. AI practitioners should make notions and concepts accessible to the general public and the impacted fields (e.g. industries, law, education). It is a truism that only law experts would have the potential to estimate IA impacts on judicial system. However, questions remain on how to connect different kind of expertise and what is the appropriate level of detail required for the knowledge exchanges. And the same consideration is true for dissemination towards society. Ultimately, society will live with decisions made by the "experts". It sounds wise to involve society in the decision process rather than risking to pay consequences later. Therefore, society also needs the key concepts to understand AI impact on their life. This was the purpose of the trial of an IA that took place in October 2018 at the Court of Appeal of Paris: gathering experts from various fields to expose challenges in law and science towards a general public.

# Introduction

The Artificial Intelligence boom is mainly due to successes of the applied mathematics field called "*Machine Learning*" (ML). The most spectacular example is the rapid progress in computer vision[1]. The capacity of computers to recognize objects, people, or vehicles are opening new horizons for technology and services such as autonomous vehicles. This radical performance shift comes from a complete change in thinking about algorithms, bringing new challenges in science (how much data do I need?), in industry (how to certify this new method?), in law (who is endorsing decisions made by an algorithm?), for society (why algorithms have made **this** decision about my life?).

# Beyond myths and magic of Artificial Intelligence

A computer can recognize a cat in a picture following the two proposed approaches. In a traditional approach, an engineer expresses rules defining what is a cat. For example, considering a picture of a cat facing the camera, a set of rules could be: this picture contains a cat if there is a brown shape having the silhouette of two triangles (cat's ears) attached to an ellipse (cat's head). Obviously, the method fails badly if the cat is not facing the camera or if it has black hair. Engineer has to create thousands of sets of rules to improve recognition performances: a cat from above, a cat sleeping with their paws on their head, etc. In Machine Learning (ML) approaches, ML algorithms create its own rules defining what is a cat. Rules are built from a dataset containing multiple pictures. A human labeled each picture, "cat" if the picture contains a cat, "no cat" otherwise. The algorithm sees one picture as a list of digits (i.e. intensity value of each pixel from the picture) and a value: "cat" or "no cat" associated. The algorithm creates mathematical rules to link every list to its value. Now that the algorithm generated the mathematical rules from all pictures, it can apply them on a new picture to return a value ("cat" or "no cat"). This simple example illustrates why confidence in algorithm grows with the size of the dataset: the more cat with different appearance and posture, the better the algorithm can learn.

While computers now exhibit equivalent or better performances than human on specific tasks, this comes with a price: these rules are not understandable for human. It is not clear what computers could see that humans do not see, leading to the unanswered questions mentioned in the introduction and many others.

---

[1] Olga Russakovsky et al., *ImageNet Large Scale Visual Recognition Challenge*, 115 INT. J. COMPUT. VIS. 211–252 (2015), http://dx.doi.org/10.1007/s11263-015-0816-y.

There are other machine learning strategies than the one described. However, the limitations stated above remain true: the need for huge dataset, specificity of the tasks, and rules are obscure for humans. Therefore, caution is advised about the notion of intelligence, there are still significant obstacles before reaching general human intelligence. The limitations presented here are only the tip of the iceberg.

## Connecting fields of expertise

Gaps between legal requirement and technology reality is likely to limit the development of "Artificial Intelligence". Protecting the public from robots could be as simple as requiring that "*a robot should not take the risk of hurting a human*". However, the best strategy for an autonomous vehicle to respect this rule is simply not moving!

In 2018, the French association *Jurisnautes* decided to organize a trial of an IA[2] after the success of the previous one on transhumanism. During the preparation of the trial, most of our energy was focus on the scenario. The diversity of our team (judges, lawyers, legal publishers, engineers) helped to tackle our first requirement: exposing the interesting legal questions caused by a massive and tragic pile-up of autonomous vehicles. This appeared naturally from our discussions as we exchange and share our knowledge about law and technology. We finally decided that the trial should address the question of a legal entity for an IA and its responsibility[3].

The second requirement was to give a realistic vision of the year 2041, when the worst would have happened. The main challenge is that science and technology are not yet ready for the dreamed system of transportations of this trial. We had to take some educated guesses about what could be expected for this both frightening and promising future. At the same time, we had to refrain ourselves from being too creative in order to expose potential contradictions and problems of our current legal system.

The third requirement was to create a scenario fair for every party: prosecutors, civil party and defense. The sentence of the Court was not written in advance, the deliberations of the Court were not prepared before the trial, we kept this old tradition! It is only at the end of the hearings that the Court with a jury of six people withdrew to deliberate.

---

[2] Françoise Barbier-Chassaing et al., *Le procès de l'intelligence artificielle et de la voiture autonome*, DALLOZ IP/IT (2018).
[3] European Parliament 2014-2019, *Civil Law Rules on Robotics*, 2103 EUR. PARLIAM. RESOLUT. 16 FEBR. 2017 WITH RECOMM. TO COMM. CIV. LAW RULES ROBOT. 22 (2016), http://www.europarl.europa.eu/meetdocs/2014_2019/plmrep/COMMITTEES/JURI/PR/2016/11-07/1095387EN.pdf.

Finally, the fourth requirement was to encompass all previous requirements in an informative and enjoyable performance for the public.

## Sharing expertise to foster public debate

Imagining expectations of the public is an additional challenge. This experience showed that public is here to hear and to learn about what they don't know. We felt right about the first requirement: exposing the questions that have not been yet unearthed for the general public. For example, the famous tramway dilemma[4] was not a central piece of the trial. Although this dilemma is captivating, it says much more about humanity than autonomous vehicles. Moreover, echoing the "simple rule" mentioned before, one day or another, someone will be hurt by a robot, then who will be responsible? This pragmatic question was also the opportunity to question the public on a more social level: does society accept the risk to be potentially harm by a robot? Would unintentional harm from a robot be an acceptable risk?

This trial has also confirmed that the public not only wants to learn that but they are also willing to understand complex problems. However, as we go deeper in the subject complexity, our message has to be crystal clear. For that reason, each party had the opportunity to invite an AI expert during their hearing to help clarify technical points. Almost freed from this technological burden, parties could focus more on legal considerations.

From the beginning to the end, the trial was almost entirely set up as a real trial. Therefore, it could have been sometimes hard to understand what parts were pure fiction (i.e. our educated guesses making the scenario work), and what parts were real (i.e. the current difficulties). A major vector for improvement would be to find a way to make this distinction explicit without breaking the pace and vivacity of the trial.

## Conclusion

The objectives of the trial were to clarify the concept of AI for autonomous vehicles, to identify its impacts on different fields, and to empower the public to explore law and science challenges. The goal was to strengthen the resilience of society against potentially harmful use of AI, but more importantly to promote an AI in service of our society. The success of the trial has shown that i) there is a real interest from the public to understand difficult topics and ii) there may never be enough exchanges between experts and society.

---

[4] Edmond Awad et al., *The Moral Machine experiment*, 563 NATURE 59–64 (2018), http://dx.doi.org/10.1038/s41586-018-0637-6.